\documentclass[12pt]{article}

\usepackage{arxiv}
\usepackage{amsmath}
\usepackage{amsfonts}
\usepackage{amsthm}
\usepackage{graphicx}
\usepackage{paralist}
\usepackage{booktabs}
\usepackage{subcaption}
\usepackage{tabularx}
\usepackage[table]{xcolor}
\usepackage{url}
\usepackage{microtype}
\usepackage{algorithm}
\usepackage{algorithmicx}
\usepackage{algpseudocode}
\usepackage{booktabs}
\usepackage[symbol]{footmisc}

\algrenewcommand\algorithmicrequire{\textbf{Input:}}
\algrenewcommand\algorithmicensure{\textbf{Output:}}
\algrenewcomment[1]{\(\triangleright\) #1}
\newcommand{\sectionname}{Section}

\graphicspath{{figures/}}

\theoremstyle{definition}
\newtheorem{exmp}{Example}

\newcommand{\equref}[1]{Eq. (\ref{#1})}
\newcommand{\para}[1]{\noindent\textbf{#1.}}
\newcommand{\ring}{\mathbb{Z}_q[x]/(x^n+1)}

\title{TPU as Cryptographic Accelerator}

\author{
Rabimba Karanjai\thanks{University of Houston, USA. Email: rkaranjai@uh.edu} \and
Sangwon Shin\thanks{Korea University, South Korea} \and
Wujie Xiong\thanks{Kent State University, USA} \and
Xinxin Fan\thanks{IoTeX, USA} \and
Lin Chen\thanks{Texas Tech University, USA} \and
Tianwei Zhang\thanks{Nanyang Technological University, Singapore} \and
Taeweon Suh\thanks{Korea University, South Korea} \and
Weidong Shi\thanks{University of Houston, USA} \and
Veronika Kuchta\thanks{Florida Atlantic University, USA} \and
Francesco Sica\thanks{Florida Atlantic University, USA} \and
Lei Xu\thanks{Kent State University, USA}
}

\date{}

\begin{document}

\maketitle

\begin{abstract}

    Cryptographic schemes like Fully Homomorphic Encryption (FHE) and Zero-Knowledge Proofs (ZKPs), while offering powerful privacy-preserving capabilities, are often hindered by their computational complexity. Polynomial multiplication, a core operation in these schemes, is a major performance bottleneck. While algorithmic advancements and specialized hardware like GPUs and FPGAs have shown promise in accelerating these computations, the recent surge in AI accelerators (TPUs/NPUs) presents a new opportunity. This paper explores the potential of leveraging TPUs/NPUs to accelerate polynomial multiplication, thereby enhancing the performance of FHE and ZKP schemes. We present techniques to adapt polynomial multiplication to these AI-centric architectures and provide a preliminary evaluation of their effectiveness. We also discuss current limitations and outline future directions for further performance improvements, paving the way for wider adoption of advanced cryptographic tools.
\end{abstract}

\section{Introduction}
In the last decades, a large number of cryptographic tools with fascinating features have been developed, such as fully homomorphic encryption (FHE)~\cite{brakerski2012fully}, zero-knowledge proof (ZKP)~\cite{groth2016size}, and post-quantum cryptography schemes (PQC)~\cite{bos2018crystals}.
FHE allows a third party to conduct arbitrary computation (addition and multiplication) on ciphertexts and ZKP helps a \textit{prover} to convince a \textit{verifier} on the correctness of a statement without disclosing the secret involved in the statement. 
These two types of cryptographic schemes can be utilized to build many interesting security applications, such as secure AI inferencing~\cite{gilad2016cryptonets,kang2022scaling} and privacy-preserving cryptocurrency exchange~\cite{xu2020privateex}.
One major obstacle that hinders the wide adoption of these schemes is their high computation cost.
PQC schemes are designed to resist attacks with quantum computers.
They are not as expensive as FHE and ZKP, but all fall into the category of asymmetric cryptography (digital signature and public key encryption/key encapsulation mechanism) and are computation-intensive by nature.

Although these schemes are different from almost all perspectives like security features and constructions, they share one similarity in terms of computation, i.e., most of their operations involve polynomial calculations defined over specific algebra structures, which usually comprise a large portion of the computation cost of these schemes and the performance bottleneck.
To fully utilize these cryptographic tools, it is important to improve the performance of the underlying basic polynomial operations, especially multiplication.

Many works have been done on improving the polynomial multiplication from the algorithmic perspective, such as Karatsuba multiplication~\cite{karatsuba1962multiplication} and number theoretic transform (NTT) based multiplication~\cite{liang2022number}.
Implementation, especially hardware-based polynomial multiplication implementation, is another important research direction.
The two major hardware platforms that are studied are GPU and Field Programmable Gate Array (FPGA).
Both GPU and FPGA support parallel computing, which is utilized to accelerate polynomial multiplication~\cite{shivdikar2022accelerating,mert2019design}.

One emerging computation hardware that is largely ignored is the tensor processing unit (TPU), or neural processing unit (NPU). 
These two names usually refer to the same type of hardware, and we use the term TPU in the rest of the paper.
The concept of TPU was originally proposed by Google to meet the explosive computation power demand of artificial intelligence (AI) applications and has been proven to be a big success~\cite{norrie2021design}.
Many TPU designs have been proposed since then, especially for the edge computing scenario. 
Compared with GPU and FPGA, TPU has its unique advantages.
On the one hand, it is closer to Application-Specific Integrated Circuits (ASIC) and can significantly accelerate specific operations with low energy costs.
On the other hand, TPU still provides a certain level of flexibility and allows end users to program it.

While TPUs are widely used in AI~\cite{huot2019high}, their application in cryptography is largely unexplored. To the best of our knowledge, this is the first work investigating TPUs for polynomial multiplication, a core operation in FHE, ZKP, and PQC. To leverage TPUs' MACs, we convert polynomial multiplications to matrix multiplications. However, TPU's AI-centric design struggles with these matrices. We address this using a carefully chosen RNS~\cite{garner1959residue} and matrix partitioning. Our prototype on Google's cloud TPU demonstrates TPU's promise for cryptographic acceleration, with significant room for improvement given the limitations of current TPU software tools.

From an algorithmic perspective, the polynomial multiplication method considered in this work uses coefficient representation and has higher computation complexity (about $\mathcal{O}(n^2)$) than the point-value representation with FFT tricks (NTT, $\mathcal{O}(n \log n)$). 
But it is still worth investigating polynomial multiplication with coefficient representation for at least two reasons:
\begin{inparaenum}[\bfseries (i)]
    \item Coefficient representation is a more convenient way to store polynomials and the multiplication is efficient when the degree is not very high; and 
    \item NTT is only applicable for polynomials with certain features (e.g., the existence of $n$th root of unity in the finite field where the polynomial coefficients are defined and the degree is $n$). 
\end{inparaenum}


In summary, the contributions of the paper include:
\begin{inparaenum}[\bfseries (i)]
    \item We design key technologies to enable the utilization of TPU for polynomial multiplications in a variety of cryptographic schemes;
    \item We develop a prototype using existing TPU hardware and software to demonstrate the feasibility of the design; and
    \item Last but not least, we discuss the potential ways to improve the TPU hardware and corresponding algorithms to further accelerate polynomial multiplication.
\end{inparaenum}

\section{Detailed Design of TPU-based Polynomial Multiplication}\label{sec-tpu-acc}
Polynomials are used intensively in various public key cryptographic schemes mainly for two reasons:
\begin{inparaenum}[\bfseries (i)]
    \item Polynomial is related to popular hard problems used in cryptography. Many PQC and FHE schemes are built atop ring LWE problems~\cite{rosca2018ring}, and using a polynomial ring is a major way to instantiate a ring LWE instance.
    \item Polynomial is an effective way for computation representation. An effective computation representation is critical for ZKP, and many ZKP schemes use polynomials in one way or another to represent the computation (or the statement) the prover needs to prove.
\end{inparaenum}
Polynomial operations, especially polynomial multiplications, usually comprise a large portion of the computation cost of these cryptographic schemes and are the performance bottleneck.



Like a typical computation device, the TPU has four types of components: I/O, control, storage, and computation.
The major part is a Matrix Multiply Unit, which is implemented as a systolic array.
This unit can perform $256 \times 256$ multiply-and-accumulate (MAC) operations on signed or unsigned integers/floating point numbers, which are essential for matrix multiplications.
In addition to the Matrix Multiplicity Unit, there is another useful computation unit, the vector unit, which is used for general computations such as the activation function.

The dedicated and optimized circuit is usually much faster and more power-efficient than CPU and GPU on MAC operations.
According to~\cite{jouppi2018domain}, a Haswell server equipped with the original version of Google TPU can achieve 92 TOPS/s with a TDP of 861W, while the server itself can only achieve 2.6 TOPS/s with a TDP of 504W. 
After the first generation, Google has released three revised versions of TPU and the latest version is V4~\cite{jouppi2021ten}.

\subsection{Polynomials Used in Cryptography and Their Operations}\label{subsec-poly-crypt}
Polynomials used in cryptography are usually defined over a ring structure $\mathbb{Z}_q[x]/(p(x))$.
From the computation perspective, there are mainly three factors to consider:
\begin{inparaenum}[\bfseries (i)]
    \item \textit{The degree of $p(x)$}. The degree of operand polynomials and operation result polynomials are less than the degree of $p(x)$. This is an important factor that dominates the complexity of the computation (addition and multiplication) in most cases.
    \item \textit{The value of $q$}. The value of $q$ determines the range of polynomial coefficients. The size (i.e., the number of bits) and form (e.g., Hamming weight of the binary representation) of $q$ determine the complexity of each operation on polynomial coefficients (e.g., addition and multiplication). 
    \item \textit{The form of $p(x)$}. All computation results need to be reduced modulo $p(x)$, and the complexity of the modulo operation is highly related to the form of $p(x)$. Generally, the modulo operation is more complex when $p(x)$ has more non-zero coefficients.
\end{inparaenum}

In this work, we mainly consider the ring $\ring$, which is widely used in the construction of different cryptographic schemes. 
Here $n$ is a power of 2, making $x^n+1$ a cyclotomic polynomial, and $q$ is an integer and does not need to be a prime number, i.e. $\mathbb{Z}_q$ does not need to be a prime field.
Another popular ring structure used in cryptography is $\mathbb{Z}_q[x]/(x^n-1)$, but the technologies described in the rest of the paper can be easily adapted to handle the case of $\mathbb{Z}_q[x]/(x^n-1)$.

A polynomial $f(x) \in \ring$ is in the form
\begin{equation}\label{equ-poly-coef}
    f(x) = a_0x^0 + a_1x^1 + \cdots + a_{n-1}x^{n-1},   
\end{equation}
where $a_i \in \mathbb{Z}_q, i=0,\ldots, n-1$.
The two operations defined on the ring are \textit{polynomial addition} and \textit{polynomial multiplication}.
Polynomial addition is straightforward, i.e., two corresponding coefficients of two polynomials are added together and then reduced modulo $q$.
Multiplication is more complex and consumes most of the computation time. 
We review two common ways to do multiplication, which are closely related to the representation of polynomials:
\begin{inparaenum}[\bfseries (i)]
    \item \textit{Multiplication with coefficient representation.}
        Coefficient representation is a natural way to store a polynomial, as demonstrated in \equref{equ-poly-coef}.
        When two polynomials defined in $\ring$ are stored using coefficient representation, a straightforward approach is first multiplying them as if they are defined in $\mathbb{Z}[x]$, and then reducing the polynomial by taking modulo $x^n+1$ and $q$ to obtain the final result.
        One disadvantage of this approach is that the intermediate result has a degree of at most $2n-2$ with coefficients as large as $(q-1)^2$.
        We may integrate these modulo operations into the multiplication to save some time.
        The complexity of a straightforward multiplication is $O(n^2)$, and using some techniques like Karatsuba algorithm can bring down the complexity to the level of $O(n^{\log_2 3})$~\cite{karatsuba1962multiplication}.
    \item \textit{Multiplication with value representation.}
        Another common way to store a polynomial is using value representation.
        For the polynomial $f(x)$ given in \equref{equ-poly-coef} represented with coefficients, one can evaluate it at $n$ different values for $x$, and store all these $n$ pairs $[x_1, y_1=f(x_1)], [x_2, y_2=f(x_2)], \ldots, [x_n, y_n=f(x_n)]$. 
        If we always use $x_i=i \in \mathbb{Z}, i=1, 2, \ldots, n$ for evaluating a polynomial, $x_i$ can be ignored and only $y_i$s need to be stored (we assume $n < q$, which usually holds).
        These pairs (or $y_i$s) fully determine the polynomial and we can convert them back to coefficient representation using Lagrange interpolation.
        The multiplication of polynomials stored using value representation is easy, i.e., we only need to perform an element-wise multiplication.
        As most cryptography schemes use polynomials with coefficient representation, one needs to convert them to value representations first. 
        A straightforward way of conversion is evaluating a polynomial at each given value, and the complexity of each evaluation is $O(n)$. 
        Since the polynomial is evaluated at $n$ values, the overall complexity is $O(n^2)$, which is more expensive than the multiplication operation in value representation. 
        Because the multiplication result may need to be involved in other cryptographic operations, we usually need to convert it back to coefficient representation.
        This inverse operation can be done using straightforward Lagrange interpolation, and its complexity is also $O(n^2)$.
        When the two parameters $q$ and $n$ of $\ring$ meet certain requirements, the conversion and reverse conversion can leverage number theoretic transform (NTT), and the complexity is reduced to the level of $O(n\log n)$. 
\end{inparaenum}

Because of the lower computation complexity, multiplication using value representation receives a lot of attention.
However, this approach has several limitations:
\begin{inparaenum}[\bfseries (i)]
    \item \textit{Size of the problem.} The lower computation complexity does not guarantee faster execution in practice. NTT is faster only for relatively large polynomial degrees. When the cryptography schemes use polynomials with low degrees, the difference between these two multiplication methods is small, and value representation can be even slower.
    \item \textit{Restrictions on parameters of $\ring$.} The low computation complexity of multiplication using value representation relies on the applicability of NTT and requires the polynomials to be NTT-friendly (e.g., $n$ is a power of $2$ and $q \equiv 1 \mod n$). While there are some works on losing these requirements~\cite{chung2021ntt}, it is hard to get rid of them completely. These requirements restrict the parameter selection for the cryptography scheme, which is not desirable.
    \item \textit{Compatibility with the hardware processor.} Value representation-based multiplication can be implemented as software but it does not guarantee that the software can fully utilize the underlying hardware capability.
\end{inparaenum}

Considering all these factors, it is worth the effort to consider polynomial multiplication using coefficient representation.
In this work, we focus on the polynomial multiplications defined in $\ring$ using coefficient representation.

\subsection{Converting Polynomial Multiplication to Matrix Operations}\label{subsec-poly-2-mat}
TPU is a special hardware that is designed to accelerate the inference operation in a deep neural network, especially the matrix operation.
Therefore, we need to convert polynomial multiplications described in \sectionname~\ref{subsec-poly-crypt} to matrix multiplications to utilize the capability of the TPU hardware.

It is relatively easy to incorporate the reduction of modulo $x^n+1$ into the multiplication process itself by modifying one of the operand polynomials. 
Example~\ref{exmp-poly-mul-mat} gives a toy example of incorporating the modulo operation into polynomial $b(x)$.

\begin{exmp}\label{exmp-poly-mul-mat}
    Considering two degree-2 polynomials defined over $R_q = \mathbb{Z}_q[x]/(x^3+1)$, $a(x)=a_0x^0 + a_1x^1 + a_2x^2$ and $b(x)=b_0x^0 + b_1x^1 + b_2x^2$. The multiplication process can be converted to a vector-matrix multiplication operation as follows:
    \begin{align}\label{equ-example-mul}
        \begin{split}
            (a_0, a_1, a_2) & \times
            \begin{bmatrix}
                b_0     &  b_1 & b_2 \\
                -b_2    &  b_0 & b_1\\
                -b_1    & -b_2 & b_0\\
            \end{bmatrix} 
             =
            \begin{bmatrix}
                a_0b_0 - a_1b_2 - a_2b_1\\
                a_0b_1 + a_1b_0 - a_2b_2\\
                a_0b_2 + a_1b_1 + a_2b_0\\
            \end{bmatrix}.
        \end{split}
    \end{align}
    The vector on the right side of \equref{equ-example-mul} presents the multiplication result, i.e., 
    $(a_0b_0 - a_1b_2 - a_2b_1)x^0 + (a_0b_1 + a_1b_0 - a_2b_2)x^1 + (a_0b_2 + a_1b_1 + a_2b_0)x^2$.
    Note that all coefficients here need to take modulo $q$.
    The matrix on the left side of \equref{equ-example-mul} is determined by the modular polynomial $x^n + 1$ and the polynomial $b(x)$.
    \qed
\end{exmp}

When the modulo polynomial is set to $x^n + 1$, the modulo operation is simpler than a general modulus polynomial.
Without loss of generality, we assume the original multiplication result is $\sum_{i=0}^{2n-2}c_ix^i$.
For a term $c_dx^d$ in the multiplication result with a degree $d \geq n$, we subtract the coefficient $c_d$ from the $(d-n)$th term, i.e., $(c_{d-n}-c_d)x^{d-n}$, to obtain the reduced result.
Following this idea, we can convert the multiplication into matrix multiplication format with moduli polynomial $x^n + 1$ as follows:

\begin{equation}\label{equ-genereal-mat}
    (a_0, a_1, \ldots, a_{n-1}) \times
    \begin{bmatrix}
        b_0         &   b_1 &   \cdots  & b_{n-1}\\
        -b_{n-1}    &   b_0 &   \cdots  &   b_{n-2}\\
        \cdots      &   \cdots  &   \cdots  &   \cdots\\
        -b_1    &   -b_2    &   \cdots  &   b_0\\
    \end{bmatrix}.
\end{equation}
For the matrix in \equref{equ-genereal-mat}, each row (except the first one) is a right shift of all elements below the main diagonal multiplied by $-1$. 

The computation given in \equref{equ-genereal-mat} is actually a vector-matrix multiplication, and it can be easily converted to matrix-matrix multiplication when one fixed polynomial (the matrix) is multiplied with many different polynomials, where each of them is represented as a vector and all the vectors (a row) form a matrix.


In other words, multiple polynomial multiplications can be converted to a single matrix multiplication, as long as one of the operand polynomials is the same.

\subsection{Challenges of Using TPU for Polynomial Multiplication}
The way of converting polynomial multiplication to matrix operation described above is not enough for utilizing TPU for cryptographic operations.
The first challenge is the degree of the polynomials. 
For polynomials with degree $n$, the corresponding matrix dimension is $n \times n$.
The degree of the polynomials involved in a typical cryptographic scheme can be large, and the corresponding matrix cannot fit the TPU hardware for computation.
The second challenge is the size of the polynomial coefficients. 
Most TPUs are designed for machine learning tasks, and a significant amount of work has been done to reduce the size of values involved in a machine learning model.
For instance, the Google TPU supports \texttt{bf16} natively, which is a 16-bit floating point number system and has lower precision compared with the standard 32-bit floating point number, but provides similar model performance~\cite{osorio2022bf16}.
However, polynomials used in cryptography usually have relatively large coefficients, from tens of bits to a few hundred bits.
Furthermore, these coefficients are generally defined on a finite and discrete algebra structure, which is not natively consistent with the data type supported by the TPU.
\tablename~\ref{tbl-poly-parameters} summarizes some polynomial parameters used by different cryptographic schemes.
We describe the way to overcome these two challenges in the following sections.

\begin{table}
    \centering
    \caption{Common polynomial parameters for cryptographic schemes.}
    \label{tbl-poly-parameters}

    \small{
    \begin{tabular}{p{1.2in}p{0.5in}p{1.2in}}
        \toprule
        \textbf{Scheme}  &   \textbf{Polynomial degree}   &   \textbf{Polynomial coefficients size}\\
        \midrule
        \rowcolor{lightgray}
        FHE (FV, BFV, CKKS) &   $2^{10}$ to $2^{14}$    &   32 to 54 bits\\
        \hline
        PQC &      $2^{8}$ to $2^{10}$  &   $\leq$ 60 bits\\
        \hline
        \rowcolor{lightgray}
        ZKP (zkSNARK, zkSTARK) &       $2^{20}$ to $2^{21}$   & 384 to 768 bits\\
        \bottomrule
    \end{tabular}
    }
\end{table}
\vspace{-7pt}
\subsection{Handling Large Coefficients}
Classical digital computers store and process values presented using a limited number of bits.
For large values, there are two strategies to process them:
\begin{inparaenum}[\bfseries (i)]
    \item\textit{Using an approximated value to replace the original one.} With this approach, limited storage is used to store a value that is close to the real value. Floating point number schemes like \texttt{IEEE 754}~\cite{kahan1996ieee} and \texttt{bfloat16}~\cite{osorio2022bf16} fall into this category of solution. This strategy is effective for a wide range of applications where the problems do not require 100\% computation accuracy.
    \item \textit{Using multiple basic units to represent the original one.} The other one is breaking a large value into several parts so each part can be stored and processed by the hardware. Accordingly, an operation on the original large values is also decomposed into a sequence of operations on the small units, which can be handled by the hardware directly. One common example is big integer operations, where an integer with more than 32 or 64 bits is stored in multiple basic data structures, and dedicated algorithms are implemented to support the common arithmetic operations~\cite{biswas2003fast}. This strategy is widely used in cryptographic libraries to support schemes such as RSA.
\end{inparaenum}

Since cryptographic operations usually deal with accurate and discrete values, the first approach is not a good option.
Applying the second strategy directly to the method described \sectionname~\ref{subsec-poly-2-mat} is not ideal either, as each coefficient multiplication/addition needs to be decomposed into several operations, which cannot be carried out by the TPU directly. 
This will lead to more complex design and lower performance.


To overcome this challenge, we propose to utilize the residue number system (RNS)~\cite{omondi2007residue} for coefficients representation and computation.
A residue number system is defined by a set of $k$ co-prime integers called the moduli $\{ m_1, m_2, \ldots, m_k \}$.
An integer $x$ is represented in the RNS by its remainder
$[x_1, x_2, \ldots, x_k]$,
where $x_i \equiv x \mod m_i$ for every $i$.

For addition/subtraction and multiplication operations of numbers represented in an RNS, it suffices to perform the same operation on each pair of residues.
For two integers represented in RNS, i.e., $x= [x_1, x_2, \ldots, x_k]$ and $y = [y_1, y_2, \ldots, y_k]$, their product $z$ is in the form 
$z=[z_1, z_2, \ldots, z_k]$,
where $z_i = x_i \times y_i \mod m_i$ for every $i$.

To recover $z$ from $[z_1, z_2, \ldots, z_k]$, we can use the Chinese Reminder Theorem since all moduli are selected to be co-prime. 
Note that we can keep the RNS representation until there is a need to convert the result back to $\ring$.

There are two advantages of using RNS to handle large coefficients in our case:
\begin{inparaenum}[\bfseries (i)]
    \item The original computation is divided into multiple instances with small values and these new instances are independent, i.e., they can be executed in parallel.
    \item Each instance can be executed using the same framework described in \sectionname~\ref{subsec-poly-crypt}, and does not require any modification.
\end{inparaenum}

\subsection{Handling High Polynomial Degrees}\label{subsec-poly-divide}
The TPU matrix multiplication hardware has a limited size and can only handle matrices efficiently within such limitations.
While it may handle much larger matrices with the help of corresponding tool chain software, the overall performance can decrease significantly.
For a polynomial with a high degree (e.g., $2^{20}$), it is hard to store the corresponding matrix completely in TPU memory.
To overcome this challenge, we adopt the divide-and-conquer method, i.e., dividing the original matrix/vector into several smaller ones that can be processed by the TPU hardware directly, computing several intermediate results, and reconstructing the final result from these intermediate values.

\begin{figure}
    \centering
    \includegraphics[width=0.8\linewidth]{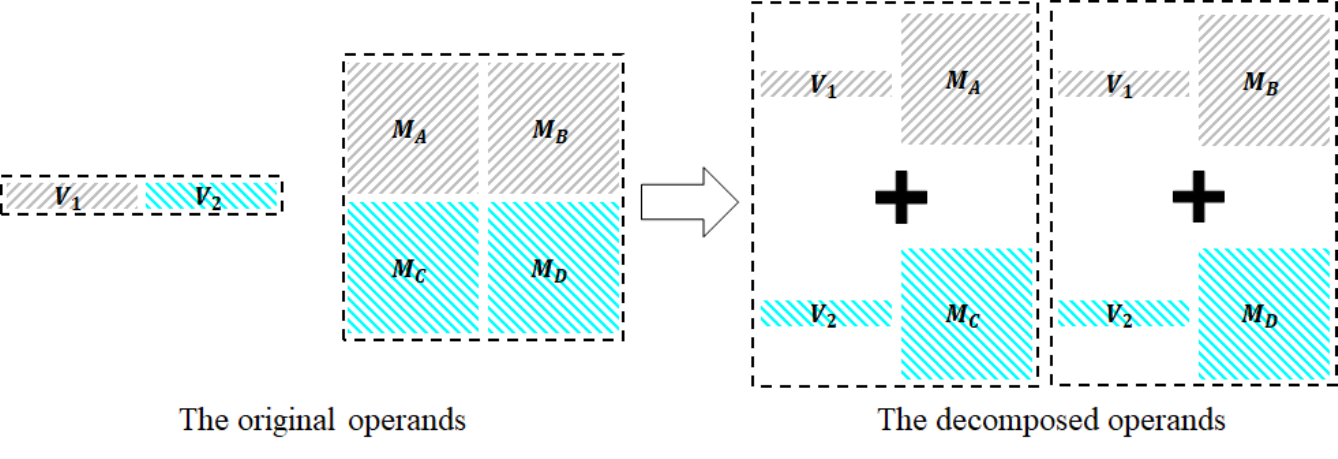}
    \vspace{-0.1in}
    \caption{Demonstration of handling of polynomials with high degree. 
    On the left side of the figure, we evenly break the vector into two sub-vectors and the matrix into four sub-matrices. On the right side, the original vector-matrix multiplication is decomposed into four vector-matrix multiplications with smaller dimensions.}
    \vspace{-0.2in}
    \label{fig-divide-large-degree}
\end{figure}

Fortunately, we can naturally divide polynomial multiplication with a large degree into several smaller instances.  
Without loss of generality, we consider the case where we need to shrink the degree of polynomials by half, i.e., instead of processing degree $n$ polynomials multiplication, we want to break it to operations on polynomials of degree $n/2$.
The vector of dimension $n$ (corresponding to one degree $n$ polynomial operand) is directly divided into two vectors of dimension $n/2$.
The matrix of dimension $n \times n$ (corresponding to the other degree $n$ polynomial operand) is divided into four sub-matrices, and each of them has dimension $n/2 \times n/2$.
\figurename~\ref{fig-divide-large-degree} depicts the way of breaking high-degree polynomials multiplication to low-degree polynomials calculations. 
To calculate and merge the intermediate results, we:
\begin{compactenum}
    \item Calculate the product of sub-vectors and sub-matrices: 
        \begin{align*}
            &r_{1A} \leftarrow V_1 \times M_A    & r_{1B} \leftarrow V_1 \times M_B    \\
            &r_{2C} \leftarrow V_2 \times M_C    & r_{2D} \leftarrow V_2 \times M_D
        \end{align*}
     All results are vectors of dimension $n/2$.
    \item Calculate the first half of the final result $r_1 = r_{1A} + r_{2C}$, which is a vector of dimension $n/2$.
    \item Calculate the second half of the final result $r_2 = r_{1B} + r_{2D}$, which is a vector of dimension $n/2$.
    \item Concatenate $r_1$ and $r_2$ to form the final result, which is a vector of dimension $n$.
\end{compactenum}

The dividing process can be repeated recursively until the vector/matrix can fit the TPU hardware. 
While this approach can reduce the dimension of vector/matrix, it also increases the number of operations.
In the above case, one single vector-matrix multiplication becomes four smaller vector-matrix multiplications with some extra operations.

While this work primarily focuses on polynomial multiplication in the ring $\mathbb{Z}_q[x]/(x^n + 1)$, the core ideas and techniques presented can be extended to other ring structures and polynomial forms commonly encountered in cryptographic applications. The conversion of polynomial multiplication to matrix multiplication, as described in Section 2.3, relies on the structure of the polynomial modulus. For different moduli, such as $x^n - 1$ or trinomials, the corresponding matrix structure will change, potentially requiring adjustments to the mapping process. However, the fundamental principle of leveraging the TPU's matrix multiplication capabilities remains applicable. Furthermore, the use of the Residue Number System (RNS) for handling large coefficients is independent of the specific ring or polynomial form. Future work will explore these generalizations in detail, including adapting the matrix conversion algorithm for different moduli and analyzing the performance impact on various cryptographic protocols with diverse polynomial structures.

When the chosen number of moduli (m) results in matrices that exceed the TPU's size limitations, we adopt a hierarchical divide-and-conquer strategy. The large matrices are further subdivided into smaller blocks that fit within the TPU's memory constraints. This involves performing multiple smaller matrix multiplications and then combining the intermediate results to obtain the final product. This approach increases the number of matrix multiplications but allows us to handle larger RNS bases and potentially improve performance by increasing parallelism. The overhead of this hierarchical division is minimized by careful scheduling and data management within the TPU environment.

\subsection{Key Parameters Selection}
For polynomial multiplication in cryptography, two important parameters are associated with large coefficients and high-degree handling.
One parameter is the size of the RNS base (and corresponding moduli), and the other parameter is the matrix dimension, which determines the size of the basic matrix calculation.

\para{RNS base selection}
Given the ring $\ring$, there are several requirements for RNS base selection.
\begin{inparaenum}[\bfseries (i)]
    \item Each modulus of the RND base should fit the TPU hardware so related calculations can be done easily.
    \item The product of all moduli of the RNS base should be larger than $q^2$ if we only need to do one multiplication operation with the given RNS system and convert it back to its original form. As we need to compute matrix multiplication, several intermediate products are added together in the computation process, so we require the product of the RNS base to be larger than $\ell q^2$, where $\ell$ is the number of addition operations.
\end{inparaenum}
At the same time, we prefer an RNS base with fewer members to reduce computation costs, so a larger modulus should be chosen first.

The selection of a suitable RNS base involves carefully considering several factors. First, the word length of the moduli is chosen to be as close as possible to the TPU's maximum integer word length, which is \textbf{8 bits} for the Google TPU v2 and v3 used in our experiments. This maximizes the utilization of the TPU's arithmetic capabilities and minimizes the overhead of splitting larger coefficients. Second, the number of moduli $(m)$ is chosen to balance the benefits of parallelism with the overhead of conversion and reconstruction. We aim for an $m$ that provides sufficient parallelism without causing excessive memory consumption or communication overhead. Third, the moduli are selected to be co-prime to ensure correct reconstruction using the Chinese Remainder Theorem (CRT). Their product must be larger than the maximum possible value of the coefficients after multiplication, which depends on the value of q and the degree of the polynomials. In our experiments, for a given m, we choose the largest possible co-prime moduli that fit within the 8-bit word length. Finally, we prioritize moduli that allow for efficient modular arithmetic on the TPU. For instance, moduli close to powers of two can simplify modulo operations and improve performance.

The value of $q$ in the polynomial ring $\mathbb{Z}_q[x]/(x^n + 1)$ and the TPU's maximum word length are crucial factors in determining the efficiency of the RNS representation and the overall performance. In our experiments, we consider values of $q$ that are larger than the 8-bit word length of the TPU. This requires representing coefficients using multiple moduli in the RNS, effectively performing multi-precision arithmetic within each modular computation. This introduces a potential bottleneck, as the TPU is optimized for single-word integer operations. However, by carefully selecting the RNS base and utilizing efficient modular arithmetic techniques, we aim to minimize the overhead of multi-precision computations. Future work will explore optimizations such as incorporating Montgomery or Barrett reduction algorithms to further improve the performance of these operations.

\para{Matrix dimension selection}
The matrix dimension selection also relies on the structure of $\ring$ and the TPU hardware parameters.
We prefer to use a larger dimension when possible as it can simplify the implementation and reduce the number of matrix multiplications.
However, a large dimension may cause two issues:
\begin{inparaenum}[\bfseries (i)]
    \item The TPU does not have enough memory to store the matrices, which can crash the application. We have observed crashes for multiple Google Cloud TPUs with large matrices.
    \item The dimension does not fit the TPU directly, and the system needs to take extra steps to break it down and the efficiency is not guaranteed. 
\end{inparaenum}


While a simple approach sets the dimension to the matrix multiplier's size, cloud TPUs often have multiple multipliers, making fine-tuning challenging. Experimentation is key to finding the optimal dimension, requiring more effort than RNS base selection.

\subsection{Putting All Pieces Together}
We have described the ways of converting polynomial multiplication to matrix multiplication, dealing with large coefficients and degrees.
Now we put these pieces together and describe the way of utilizing TPU for polynomial multiplication used in cryptographic schemes.
The algorithm consists of three major steps:
\begin{inparaenum}[\bfseries (i)]
    \item Determination of the parameters to decompose the original operand polynomials to fit into the TPU hardware;
    \item Computation of intermediate multiplication results utilizing the TPU hardware; and
    \item Reconstruction final polynomial multiplication results from the intermediate values produced by the previous step.
\end{inparaenum}
Algorithm~\ref{alg-tpu-straight-poly-mul} summarizes the process.

\begin{algorithm}
    \caption{Polynomial multiplication using TPU.}
    \label{alg-tpu-straight-poly-mul}
    \begin{algorithmic}[1]
      \Require{Two polynomials $a(x), b(x) \in \ring$, TPU parameters $\mathit{para}$}
      \Ensure{$c(x) = a(x)b(x) \in \ring$}
      
      \Statex \Comment Creating the configuration information based on the problem and the hardware
      \State $\mathit{config} \leftarrow$ \Call{GenConfig}{$(\ring, \mathit{para})$}

      \Statex \Comment Extract the list of moduli from $\mathit{config}$, which is used to build the RNS base
      \State $modList \leftarrow$ \Call{ExtractModuli}{$\mathit{config}$}

      \Statex \Comment Extract matrix dimension information from $\mathit{config}$
      \State $dimList \leftarrow$ \Call{ExtractDimensions}{$\mathit{config}$}

      \Statex \Comment Use selected parameters to convert polynomials to matrix form for multiplication on TPU
      \State $A \leftarrow$ \Call{ConvertPoly}{$a(x), modList, dimList$}
      \State $B \leftarrow$ \Call{ConvertPoly}{$b(x), modList, dimList$}

      \Statex \Comment Calculate matrix multiplication using TPU
      \State $C \leftarrow $ \Call{TPUMul}{$A, B$}

      \Statex \Comment If needed, convert back the multiplication result to coefficient representation
      \State $c(x) \leftarrow$ \Call{ConvertMatrix}{$C, modList, dimList$}

      \State \Return $c(x)$
    \end{algorithmic}
  \end{algorithm}

\section{Experiments and Evaluation}\label{sec-experiments}
Due to the rising popularity of AI applications~\cite{karanjai2024solmover}, there is a growing interest in the design and development of TPU hardware~\cite{olsen2017proposal}. However, the number of TPU products accessible to general developers remains limited. One of the most widely used options is Google's TPU, which is what we primarily evaluate our design using.

\para{Overview of Google Cloud-based TPU} Google offers two primary methods for accessing its TPU hardware: creating a TPU-equipped virtual machine (VM) in the cloud or utilizing Colab, an online development platform with TPU connectivity. Both approaches support popular frameworks such as TensorFlow, PyTorch, and JAX (\url{https://github.com/google/jax}), which are all Python-based and originally designed for AI applications.

For our prototyping and benchmarking, we chose the TPU VM paired with JAX, as it provides the greatest flexibility. This setup allows us to configure the TPU VM with different hardware versions and topologies. JAX also supports non-AI computations and includes tools for optimizing program performance. Currently, Google Colab only allows requesting a TPU instance without additional hardware configuration options. Furthermore, Colab does not yet support JAX 0.4, which we used in our experiments.
\begin{figure*}[h]
    \begin{subfigure}[t]{0.9\textwidth}
        \centering
        \includegraphics[width=1.37in]{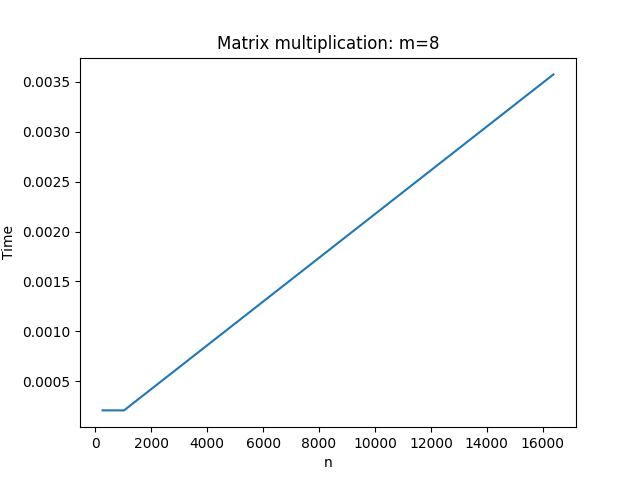} 
        \includegraphics[width=1.37in]{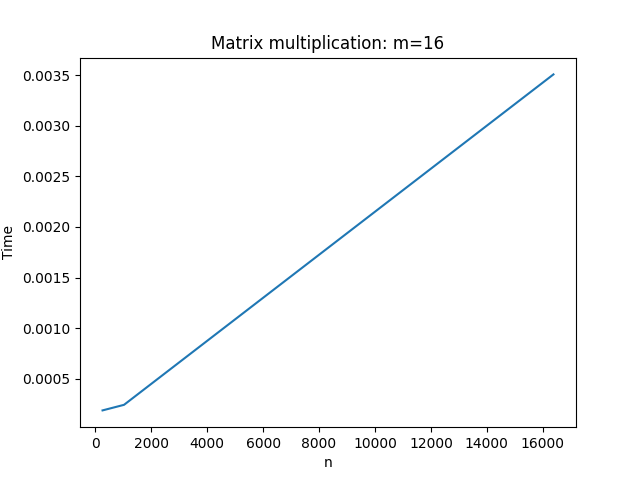}
        \includegraphics[width=1.37in]{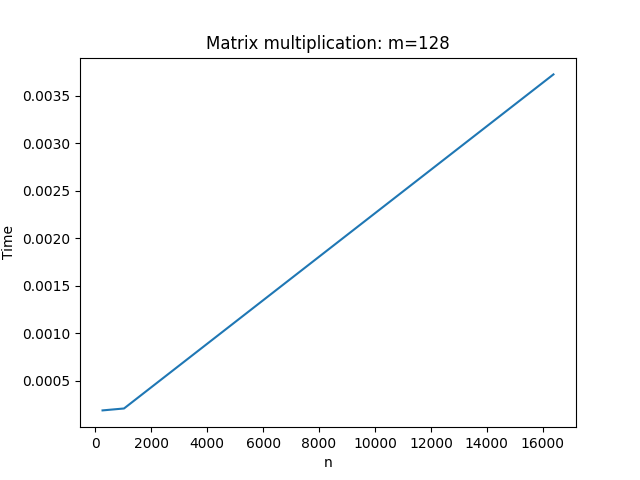}
        \vspace{-0.1in}
        \caption{TPU v2 with $m=8, 16, 128$}
        \label{fig-tpu-v2-perf}
    \end{subfigure}
    \begin{subfigure}[t]{0.9\textwidth}
        \centering
        \includegraphics[width=1.37in]{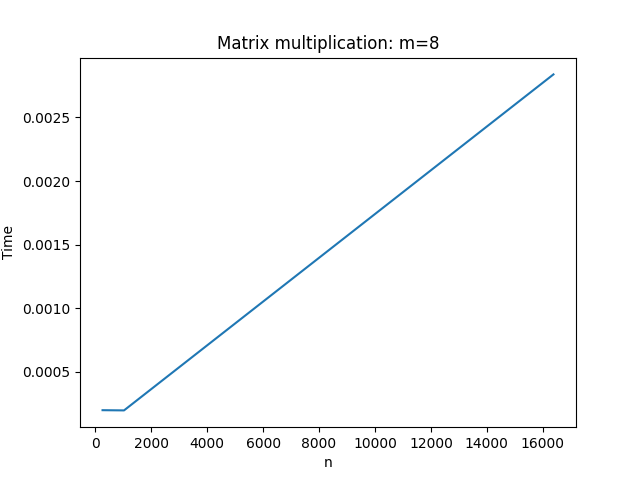}
        \includegraphics[width=1.37in]{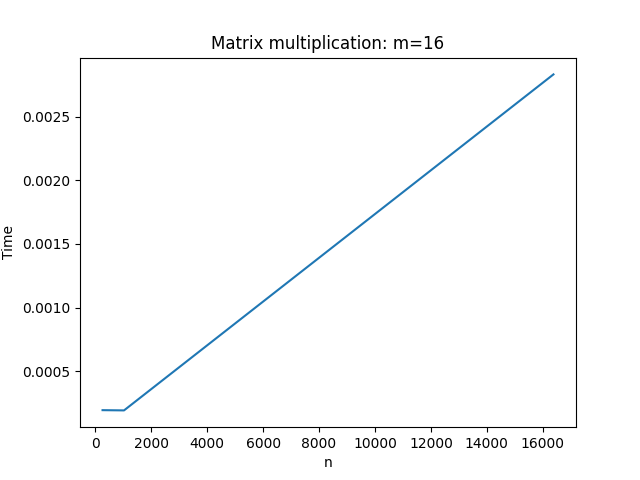}
        \includegraphics[width=1.37in]{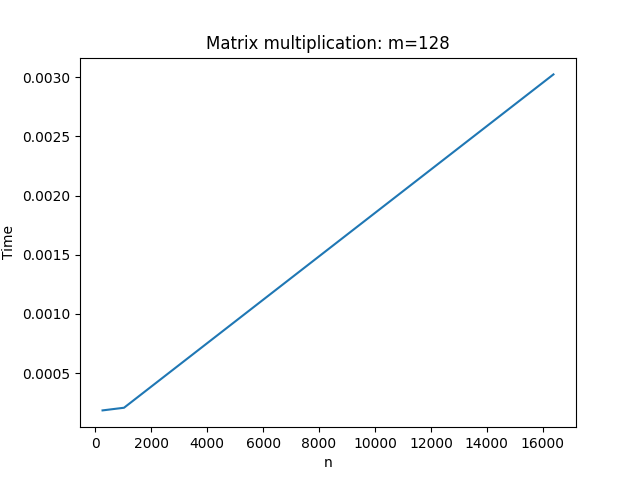}
        \vspace{-0.1in}
        \caption{TPU v3 with $m=8, 16, 128$}
        \label{fig-tpu-v3-perf}
    \end{subfigure}
    \vspace{-0.15in}
    \caption{Summary of experiment results using Google TPU with different configurations.}
\end{figure*}

\para{Prototype and evaluation}
The most computation-intensive part of polynomial multiplication in $\ring$ is the matrix multiplication operation.
As we discussed in Section~\ref{sec-tpu-acc}, the size of coefficients and the degree of the polynomial will affect the dimensions of the matrix we have to deal with.
We evaluate the performance using three polynomial degrees, i.e., $n=2^8, 2^{10}, 2^{14}$.
For the RNS base size, we also consider three values, i.e., $m=8, 16, 128$.
\tablename~\ref{tbl-env-description} describes the two benchmarking environments.

\begin{table}
    \centering
    \caption{Benchmarking environments on Google Cloud.}
    \label{tbl-env-description}
    \small{
    \vspace{-0.1in}
    \begin{tabular}{p{0.8in}p{2.2in}}
        \toprule
        \textbf{Env} &  \textbf{Description (Physical Cores)}\\
        \midrule
        Env 1 (\figurename~\ref{fig-tpu-v2-perf})   &   Ubuntu 22.04, TPU v2-8: 8 cores, 64 GB memory.\\
        Env 2 (\figurename~\ref{fig-tpu-v3-perf})   &   Ubuntu 22.04, TPU v3-8: 8 cores, 128 GB memory.\\
        \bottomrule
    \end{tabular}
    \vspace{-0.2in}
    }
\end{table}

\figurename~\ref{fig-tpu-v2-perf} and \figurename~\ref{fig-tpu-v3-perf} summarize the benchmarking results using two execution environments Env 1 and Env 2 respectively.
For each figure, we fix the value of $m$ and the x-axis is different values of $n$. 
The y-axis is the execution time and the unit is the second.
We observe several things from the experiment results:
\begin{inparaenum}[\bfseries (i)]
    \item For small matrix dimensions, the computation time is a constant value. After the threshold, the computation time increases linearly with the value of $n$. Note that the overall computation complexity is $O(m \times n \times n)$ (if the computation is done using the straightforward algorithm), which is not consistent with the experiment results. One potential reason is that the TPU architecture carries out the computation in a highly parallel manner and reduces the latency.
    \item The slopes of curves for different $m$ values are similar, which means that $m$ does not affect the cost much. The potential reason is that the TPU hardware supports $128 \times 128$ matrix multiplication, and the largest $m$ we choose is within this range.
    \item TPU v3 is faster than TPU v2, but the difference is not significant, especially for cases with small parameters. This may be because both versions have the same number of matrix multiplication hardware (\tablename~\ref{tbl-env-description}).
\end{inparaenum}

\para{Limitations of the current benchmarking}

The experimental results demonstrate that TPUs hold significant promise as cryptography accelerators. However, the primary limitation of our current benchmarking lies in the inability to control execution at a granular level, potentially leaving TPU capabilities underutilized. For instance, while TPU v3 offers substantially larger memory, the current programming environment does not provide direct control over its usage. When multiple matrix multiplications are required, creating a pipeline could reduce overall latency. Unfortunately, Google's TPU framework does not yet support such fine-tuning.

\section{Further Improving Polynomial Multiplication with TPU}\label{sec-improve-coef-mul}
In the previous description, we focus on solving the feasibility challenges of leveraging TPU to accelerate polynomial multiplication, which is the performance bottleneck of various cryptographic schemes.
In this section, we describe some techniques to further optimize the system.

\para{Karatsuba multiplication}
The idea of Karatsuba multiplication~\cite{karatsuba1962multiplication} is trading one expensive multiplication with multiple cheap addition/subtraction operations.
For two polynomials $f(x)$ and $g(x)$ with degree $n$, we can rewrite them as
$f(x) = f_1(x) x^m + f_0(x)$ and $g(x) = g_1(x) x^m + g_0(x)$,
where $f_1, g_1$ are polynomials of degree $n-m$, and $f_0, g_0$ are polynomials of degree $m-1$.
The product is then 
\begin{align*}
    f(x)g(x) &= (f_1(x) x^m + f_0(x))(g_1(x) x^m + g_0(x))\\
             &= h_2(x)x^{2m} + h_1(x)x^m + h_0(x),
\end{align*}
where
$h_2(x)  = f_1(x)g_1(x),
 h_1(x)  = (f_1(x)+f_0(x))(g_1(x)+g_0(x)) - h_2(x) - h_0(x),
 h_0(x)  = f_0(x)g_0(x)
$.

This technique can also be applied to accelerate TPU-based polynomial multiplication. Specifically, the original polynomials are decomposed into multiple lower-degree polynomials, and the TPU is then employed to perform the multiplications. This method can be used in place of the high-degree polynomial handling approach described in \sectionname~\ref{subsec-poly-divide}. However, a key challenge lies in the TPU's inefficiency for polynomial addition and subtraction. The use of Karatsuba multiplication may require additional hardware, introducing communication overhead and complicating optimization.

\para{Fine-tuning the design} In this work, we primarily address the feasibility of using TPUs for polynomial multiplication. In practice, cryptographic schemes often require multiple polynomial multiplications in a specific sequence. Several strategies can be considered to fine-tune the computation: \begin{inparaenum}[\bfseries (i)] \item \textit{Fixing one operand polynomial as much as possible.} One operand polynomial needs to be converted to matrix form to incorporate the modulo operation into the multiplication. Creating the matrix and loading it into TPU memory is costly compared to matrix multiplication itself. Therefore, reusing a pre-generated and loaded matrix minimizes the preparation overhead. \item \textit{Optimizing computation and I/O.} Polynomial multiplication in cryptography is both computation- and I/O-intensive, and the environment is complex. As TPUs function as passive accelerators, data must be transferred from the host machine to the TPU card. Since a machine typically has multiple TPUs, selecting the appropriate TPU at a given time is crucial. Moreover, intermediate results generated by the TPU might need to be sent back to the host for further processing, so ensuring the TPU remains occupied with other tasks during this transfer is essential. \end{inparaenum}


\vspace{-7pt}
\section{Related Works}
Accelerating multiplication of polynomials defined in specific rings has been studied from different perspectives.

\para{Algorithmic approaches}
Multiplication acceleration methods are studied for both coefficient and value representations, e.g., Karatsuba algorithm~\cite{karatsuba1962multiplication} and various NTT approaches~\cite{chung2021ntt}.
Some works also consider more efficient modulo operation, which is heavily involved in polynomial operations~\cite{posch1995modulo}. 
These works are orthogonal to our work and have the potential to be utilized to further improve the polynomial multiplication performance with the TPU.

\para{Hardware-oriented optimization approach}
This line of research investigates the utilization of existing hardware features to accelerate polynomial multiplication on specific rings.
Several types of hardware are studied for this purpose, including CPU~\cite{aguilar2016nfllib}, GPU~\cite{ni2023enabling}, and FPGA~\cite{mert2020fpga}.
Compared with other types of hardware, TPU provides a new balance between flexibility, power consumption, and parallelism.
\tablename~\ref{tab-cmp-hardware} compares TPU with other computation platforms.

\begin{table}[ht]
    \caption{Comparison of major computation platforms.}
    \label{tab-cmp-hardware}
    \centering
    {\small
    \begin{tabular}{p{0.5in}p{0.6in}p{1.1in}p{0.5in}}
        \toprule
        \textbf{Hardware}    & \textbf{Flexibility}   &   \textbf{Power Consumption}    & \textbf{Parallelism}\\
        \midrule
        CPU     &   Very High   &   High        &   Low   \\
        GPU     &   High        &   Very High   &   High\\
        FPGA    &   Moderate    &   Low         &   High\\
        TPU     &   Low         &   Low         &   High\\
        \bottomrule
    \end{tabular}
    }
\end{table}

\para{TPU related works}
Most existing works on TPU focus on its architecture~\cite{jouppi2018domain,jouppi2023tpu} and applications in AI~\cite{kumar2019scale,kimm2021performance}, with a few exceptions considering TPU for non-AI use cases.
For instance, Hsu and Tseng developed a tool to use edge TPU for general-purpose computation~\cite{hsu2021accelerating}, and Lewis et.al discussed the use of TPU for large-scale linear algebra computation~\cite{lewis2022large}.
None of these works consider the special requirements of cryptographic schemes.

\section{Conclusion and Future Work}\label{sec-conclusion}

Accelerating polynomial multiplication within specific algebraic structures is essential for enhancing the performance of various cryptographic schemes. In this work, we explore the use of Tensor Processing Units (TPUs), to accelerate polynomial multiplication and, consequently, cryptographic operations. We address two primary challenges in this endeavor: (i) employing the Residue Number System (RNS) to manage large coefficients unsupported by TPUs, and (ii) utilizing a divide-and-conquer approach to handle high-degree polynomials that exceed hardware capacity. 

For future work, we intend to focus on the following three key areas: \begin{inparaenum}[\bfseries (i)] \item Explore the application of Tensor Processing Units (TPUs) for NTT-based polynomial multiplication; \item Enhance the current prototype by optimizing the hardware configuration tailored to specific polynomial rings; and \item Design and validate end-to-end cryptographic schemes utilizing TPUs, while investigating potential optimization strategies. \end{inparaenum}

\section*{Acknowledgments}
This research was supported by NATO SPS G7200, NCAE H98230-24-1-0097. Research also supported with Cloud TPUs from Google's TPU Research Cloud (TRC) and Google Developer Expert program. 
We thank these organizations for their funding and support.

\bibliographystyle{unsrtnat}
\bibliography{ref}

\end{document}